# FREQUENCY AND TIME DOMAIN PACKET SCHEDULING BASED ON CHANNEL PREDICTION WITH IMPERFECT CQI IN LTE

Yongxin Wang[1], Kumbesan Sandrasegaran[2], Xinning Zhu[3], Jingjing Fei[4] Xiaoying Kong[5] and Cheng-Chung Lin [6]

[1] FEIT, University of Technology, Sydney, Australia
Yongxin.Wang-1@student.uts.edu
[2] FEIT, University of Technology, Sydney, Australia
Kumbesan.Sandrasegaran@uts.edu.au
[3]Beijing University of Post and Telecommunications, Beijing, China
zhuxn@bupt.edu.cn
[4] CSE, University of New South Wales, Sydney, Australia
jingjingf@cse.unsw.edu.au
[5] FEIT, University of Technology, Sydney, Australia
Xiaoying.Kong@uts.edu.au
[6] FEIT, University of Technology, Sydney, Australia
Cheng-Chung.Lin@eng.uts.edu.au

**ABSTRACT**

*Channel-dependent scheduling of transmission of data packets in a wireless system is based on measurement and feedback of the channel quality. To alleviate the performance degradation due to simultaneous multiple imperfect channel quality information (CQI), a simple and efficient packet scheduling (PS) algorithm is developed in downlink LTE system for real time traffic. A frequency domain channel predictor based on Kalman filter is first developed to restore the true CQI from erroneous channel quality feedback. Then, a time domain grouping technique employing the joint of Proportional Fair (PF) and Modified Largest Weighted Delay First (M-LWDF) algorithms is used. It was proved this proposal achieves better performance in terms of system throughput and packet loss ratio by simulation results.*

## 1. INTRODUCTION

To provide a long-term development of the third generation (3G) services and better quality of wireless communication in terms of high performance and capacity, simplicity and wide range of terminals, "Super 3G"[1]. standard that expands upon the 3rd generation partnership project (3GPP) specification was proposed in 2004, and it was called long-term evolution (LTE) within 3GPP. Orthogonal frequency division multiple access (OFDMA) technology was chosen in the downlink LTE instead of other access technologies (e.g. Frequency Division Multiple Access - FDMA, Time Division Multiple Access - TDMA, and Code Division Multiple Access - CDMA) due to its robustness to inter-symbol interference (ISI) and its immunity to frequency-selective fading nature of the mobile channels[2].The use of the OFDMA is best suited for simultaneous support for multiple high data rate real time applications.

In downlink LTE, packet scheduling is the term used when a packet scheduler select a user, based on a priority such as good channel quality and packet delay information, to transmit its packets on the radio resources. The LTE standard defines a resource allocation structure in both time and frequency domains. The minimum downlink LTE transmission unit that can be allocated to a user is called a physical resource block (PRB) which includes two resource blocks (RB). In the frequency domain (FD), a PRB consists of 12 consecutive sub-carries of 180 kHz total bandwidth and is of 1 ms duration in time domain (TD). This feature encourages the use of both time and frequency domain packet scheduling algorithms.

There are many well-known packet scheduling algorithms designed in time domain, such as PF algorithm[3] and M-LWDF algorithm[4]. The PF scheduler provides a good trade-off between throughput maximization and fairness guarantee by trying to serve a user whose channel quality achieves its peak maximum to average ratio in any given scheduling interval. However, because of a lack of consideration of the buffer information of each user, such as the delay of the head of line (HoL, the HoL packet of a user is the packet that has resided the longest in its buffer at the enhanced Node B (eNB)), the PF scheduling algorithm is not expected to support real time services. The M-LWDF algorithm was developed to support real time traffic in code division multiple access-high data rate (CDMA-HDR) system. This algorithm is supposed to have a reasonably overall performance due to the consideration of packet delay information, the instantaneous data rate and average throughput. The adaptation of these two packet scheduling algorithms into frequency domain for downlink LTE can be found in[5].

Various frequency domain packet scheduling algorithms that are able to allocate resources to multiple users in one transmission time interval (TTI) have been proposed. These algorithms calculate the channel quality of each user on every PRB, and select a user with the highest priority on that PRB. Recent FDPS study in [6] has shown that FDPS can provide around 40% improvement in both average system capacity and cell throughput over time domain only scheduling. In[7-10], algorithms that are based on resource allocation and resource assignment techniques were developed. The resource allocation technique calculates the number of RBs that is to be allocated to each user in each TTI. If the PRBs are insufficient to satisfy all the users' allocation requirement, the priority of each user will be calculated to decide who can be selected to obtain the competitive resource. Thereafter, the resource assignment technique allocates the PRBs to the user who has the highest priority for that PRB based on a scheduling algorithm. These resource allocation-assignment scheduling algorithms are quite complex due to the fact that it involves many steps on the RB basis, resulting in a large number of iterations. .

To balance the computational complexity and scheduling efficiency, joint time frequency schedulers are investigated in[11-13]. These schedulers operate in two steps, where the first step performs a pre-selection based on a TD scheduling algorithm to limit the number of multiplexed users that go into the frequency scheduler. The second step consists of a resource allocation to users from the TD scheduler according to a frequency scheduling algorithm. The computational complexity can be reduced by the employment of the TD scheduler which either does not consider each user's channel quality or only take the channel quality across the whole bandwidth into consideration. Furthermore, the algorithm can benefit from the limitation of the number of input users to the FD scheduler. In fact, these algorithms are still complex due to the simultaneous employment of two packet schedulers.

However, all the above mentioned packet scheduling algorithms assume that the measured CQI of each user are ideally fed back to the eNB without any error, which is impossible in a practical wireless channel. In the downlink LTE system, feedback and processing delays cause a mismatch between the current channel state and the CQI received by the scheduler[14], along with interference, multi-path fading, and shadowing. These might cause severe performance degradation especially for a channel-dependent packet scheduler that highly relies on an accurate CQI report. In [15], performance of adaptive OFDM systems due to outdated CQI was improved by channel prediction while [16] studied the effect of multiple estimations to mitigate the degradation due to CQI delay as well as OFDM channel estimation error. The problem caused by imperfect CQI in OFDMA downlink system has been studied in[17,18]. In [17], optimal resource allocation algorithms employing a dual optimization framework were developed, in which the complexity for discrete rate scenario was shown by simulation results to be much less than continuous rate scenario. In [18], a delay-sensitive cross-layer design framework with heterogeneous delay requirements was proposed so as to guarantee a fixed target outage probability in optimal power and subcarrier allocation under heterogeneous users' delay constraints

Generally, imperfect channel conditions can lead to three types of inaccurate CQI: erroneous CQI, outdated CQI and unavailable CQI when the delay of the outdated CQI is longer than the scheduling interval. Outdated CQI is the most common situation due to the fact that feedback and processing delay lead to a mismatch between the current channel state and the received CQI. The effect of outdated CQI has been analysed in [14, 19, 20]. In [14], several CQI estimators with CQI delay were studied and numerical results showed that the proposed prediction method based on Stochastic Approximation had the best performance and a lower computational complexity. In [19], the Probability Distribution Function (PDF) of the CQI corresponding to an predicted CQI was proved to improve the Multi carrier Proportional Fair (MPF) algorithm in terms of system throughput in downlink OFDMA systems. In [20], outdated and unavailable CQI scenarios were simulated by sporadic traffic. Results showed that the proposed resource allocation scheme based on an optimal HARQ and adaptive modulation and coding policy improves scheduling efficiency for LTE-advanced networks with delay constraint. In [3], simulation results presented that the improvement method can alleviate the detrimental effect caused by CQI delay.

Several multi-user systems predictors were investigated in [21]. Wiener filter for iterative channel estimation in OFDMA systems, channel estimation performance without memory (Block-Least Squares Estimation, Block-LSE), and channel estimation with memory employing Kalman filter are discussed in [22], [23], and [24], respectively.

Among the existing packet scheduling algorithms in LTE downlink system, only limited number of them have an overall acceptable performance under simultaneous multiple imperfect channel states. However, due to the large-scale computational problem, the practical implementation of the algorithms is difficult. The aim of this paper is to propose an efficient and simple packet scheduling algorithm that provides good system performance when there are simultaneous multiple CQI problems. This scheduling algorithm involves a FD predictor based on Kalman filter and a simple TD grouping technique which allows alternate use of two different FD schedulers. Simulation results show that the packet scheduling is optimized with the combination techniques, which results in a performance improvement.

The rest of this paper is organized as follows: The downlink LTE system model is provided in Section 2. The FD channel quality predictor based on Kalman filter is described in Section 3 followed by introduction of the TD grouping technique in Section 4. Section 5 gives an overview of the proposed algorithm. Simulation results are presented in Section 6. Finally, Section 7 concludes the paper.

## 2. DOWNLINK LTE SYSTEM MODEL

The LTE deploys a simple architecture that only consists of eNBs which perform packet scheduling along with other radio resource management (RRM) functions. In this paper, 7

International Journal of Wireless & Mobile Networks (IJWMN) Vol. 5, No. 4, August 2013

hexagonal cells of 5 MHz bandwidth with 25 PRBs and 2 GHz carrier frequency is modelled. Each PRB is of 1 ms duration and contains 14 OFDMA symbols and a total of 168 resource elements (REs) with the usage of a normal cyclic prefix (CP). Table 1 gives the main downlink LTE system parameters.

It is assumed in this paper, each eNB is located at the centre of each cell with 100 meter radius. Users are uniformly distributed within each cell and inter-cell interference remains constant. Each user moves within the cell at a constant speed in a constant direction which is initialized as a random value and is wrapped-around whenever it reaches the cell boundary [25]. Users only have real time applications.

Users compute their instantaneous channel quality (e.g. signal-to-noise-ratio, SNR) on each RB which is mapped into a CQI, and report it to the serving eNB at each TTI. The reported CQI value is then used to make a scheduling decision and determine the data rate of a user for packet transmission. In this paper, it is assumed that all the CQI reports have 3 ms delay [26] and become unavailable at regular intervals (i.e. the CQIs are set to 0 every 10 ms). Equation (1) gives the mathematical expression of the instantaneous SINR experienced by user *i* on RB *j* at time *t*:

$$r_{i,j}(t) = Efficiency_{i,j}(t) * \frac{RE_{data}}{TTI} \qquad (1)$$

where $Efficiency_{i,j}(t)$ is the efficiency (in bits/RE) of PRB *j* of user *i* at time *t*. $RE_{data}$ is the total number of Res specified for downlink data transmission.

Table 1. Main downlink LTE system parameters

| System Parameters | Values |
|---|---|
| Cellular layout | 7 hexagonal cells |
| Bandwidth | 5 MHz |
| Carrier frequency | 2 GHz |
| Mode of operation | FDD |
| Number of RBs | 25 |
| Number of sub-carriers per RB | 12 |
| Total number of Sub-carriers | 300 |
| Sub-carrier spacing | 15 kHz |
| Scheduling interval (TTI) | 1 ms |
| Number of OFDMA symbols per TTI | 14 (Normal CP) |
| Total number of REs | 168 |
| Total eNB transmit power | 43.01 dBm |
| Erroneous CQI | Delay and unavailable |
| User speed | 120km/h |

## 3. FD CHANNEL QUALITY PREDICTOR BASED ON KALMAN FILTER

In 1960, Kalman filter was proposed as an efficient recursive solution of the least-squares method in [24]. The Kalman filter is the best possible (optimal) estimator for a wide variety of problems, due to its simple and clear mathematical model with the robustness and efficiency in the resolution of various problems.

## 3.1. Prediction Process

As an optimal autoregressive prediction method, Kalman filter has been applied to channel condition prediction and it enables to forecast current channel state from the state in the last time slot. Specially, the classical Kalman filter method includes two steps: estimation and correction. In the estimation step, the current CQI value is estimated from the CQI in last TTI while the estimated current CQI conditioned on last TTI is improved according to the observed CQI value that is received from each user's CQI report in the correction step. The equations are given below:
Estimation step:

$$X(t|t-1) = \Phi X(t-1|t-1) \qquad (2)$$

$$P(t|t-1) = \Phi P(t-1|t-1)\Phi' + Q \qquad (3)$$

where $X(t|t-1)$ is the estimate of *X(t)* based on the time *t-1*. $\Phi$ is transition matrix that connects two consecutive states, *X(t)* and *X(t-1)*, of the system, $P(t|t-1)$ is the covariance of $X(t|t-1)$, and $P(t-1|t-1)$ is the covariance of $X(t-1|t-1)$. The random variable $Q$ denotes noise error matrix. The aim of calculating $P(t|t-1)$ is to estimate the accuracy of $X(t|t-1)$. If $P(t|t-1) \to 0$, which means the estimate $X(t|t-1)$ is so close to the realistic CQI value that only limited correction need to be made, otherwise, if $P(t|t-1) \to 1$, which means the estimate $X(t|t-1)$ is considered to be unreliable, and it will reduce to the observation.

Correction step:

$$Z(t) = HX(t) + R \qquad (4)$$

$$K(t) = P(t|t-1)H^T[HP(t|t-1)H^T + R]^{-1} \qquad (5)$$

$$\tilde{X}(t) = \tilde{X}(t|t-1) + K(t)[Z(t) - H\tilde{X}(t|t-1)] \qquad (6)$$

$$P(t) = [I - K(t)H]P(t|t-1) \qquad (7)$$

Equation (4) calculates the observed value $Z(t)$ from the realistic channel, where *H* is the channel gain, *R* is observed error matrix. In our situation, Z(t) can be obtained directly from the CQI value reported by the users.

Equation (5) gives the Kalman gain $K(t)$, where the superscript *T* represents transpose. Kalman gain can be regarded as a weight factor that "decides" the amount of improvement to the estimate $X(t|t-1)$

From the Equation (2) - (5), we can get the optimal estimation of current state $\tilde{X}(t)$ from Equation (6), which is an update from $X(t|t-1)$.

In addition, we need to update the covariance to use in the next packet scheduling interval, and this process is given by Equation (7), where *I* is the identity matrix.

The procedure is shown in Figure. 1.

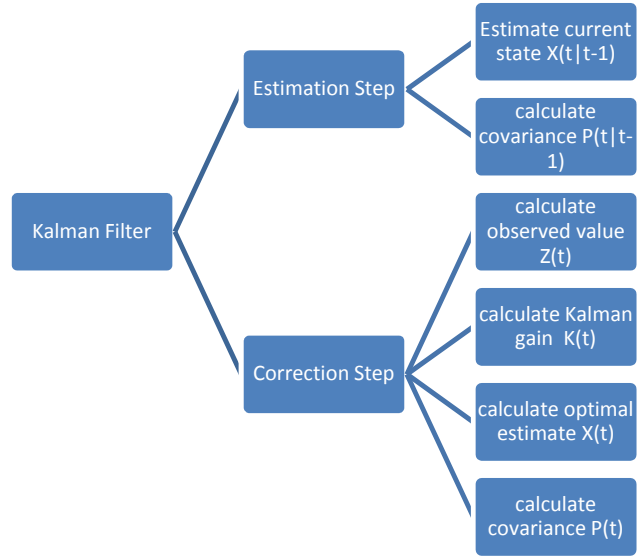

Figure 1: the procedure of Kalman filter

### 3.2. Initialization of Kalman Filter

Although few papers and books talk about initialization of Kalman filter, it is rather important to mention that if either the matrix Q or R are disarranged once, they keep disarranged and will not be updated and the Kalman process might give bad predictions [27].

At first, the expressions of state matrix *X(t)* and observed matrix *Z(t)* can be derived from the Taylor series:

$$x(t) = x(0) + \dot{x}(0)t + \frac{1}{2}\ddot{x}(0)t^2 + \cdots \qquad (8)$$

where $\dot{x}(0)$ indicates the first derivative of **x** at initial time *t=0*, $\ddot{x}(0)$ indicates the second derivative of **x** at initial time *t=0* etc. The Taylor series is generally reduced to:

$$x(t) = x(0) + \dot{x}(0)t + \frac{1}{2}\ddot{x}(0)t^2 \qquad (9)$$

In channel condition prediction, x(0) is the Signal-to-Interference-plus-Noise-Ratio (SINR) of user *i* on RB *j* at scheduling interval 0, denoted by $\gamma_{ij}$, we can consider $\dot{x}(0)$ as the change rate of SINR, represented as $v_i$ and $\ddot{x}(0)$ is the gradient of $v_i$, denoted by $b_i$. $\gamma'_{ij}, v'_i, b'_i$ are the observed $\gamma_{ij}, v_i, b_i$, respectively. On this basis, the state equation and observation equation can be respectively established as follows, where *T* denotes the transpose.

$$X(t) = (\gamma_{ij}, v_i, b_i)^T$$
$$Z(t) = (\gamma'_{ij}, v'_i, b'_i)^T$$

Then, the transition matrix can be determined as:

$$\Phi = \begin{vmatrix} 1 & t & t^2/2 \\ 0 & 1 & t \\ 0 & 0 & 1 \end{vmatrix}$$

Thereafter, since the observed matrix is directly obtained from the state matrix, the channel gain matrix can be determined as the identity matrix:

$$H = \begin{vmatrix} 1 & 0 & 0 \\ 0 & 1 & 0 \\ 0 & 0 & 1 \end{vmatrix}$$

After the determination of state matrix, the observed matrix, the transition matrix and the channel gain matrix, covariance matrix of the measurement error $R$ can be initialized Let $X_{e,x}(t)$, $X_{e,y}(t)$, $X_{e,z}(t)$ denote the random variable that describes the measurement error. The error measurement covariance is

$$R = E[(X_{e,x}(t), X_{e,y}(t), X_{e,z}(t))^T (X_{e,x}(t), X_{e,y}(t), X_{e,z}(t))] \tag{10}$$

The elements of the principal diagonal are the covariance of $X_{e,x}(t), X_{e,y}(t), X_{e,z}(t)$ respectively. Other elements are the covariance of each two variables, since each variable is completely independent. The matrix is:

$$R = \begin{bmatrix} \sigma_{e,x}^2 & 0 & 0 \\ 0 & \sigma_{e,y}^2 & 0 \\ 0 & 0 & \sigma_{e,z}^2 \end{bmatrix}$$

Finally, the state matrix $X(t-1|t-1)$ can be simply set to zero. However, it is obvious that it is not zero, hence, we get an error in the first estimation and we can set the covariance matrix of the estimation error $P(t-1|t-1)$ as:

$$P_0 = \begin{bmatrix} \gamma_{ij}^2 & 0 & 0 \\ 0 & v_i^2 & 0 \\ 0 & 0 & b_i^2 \end{bmatrix}$$

## 4. TD GROUPING TECHNIQUE

Papers [13, 28] proved that packet scheduler can benefit from the use of two scheduling algorithms. However, the simultaneous use of two different scheduling algorithms can lead to an increase in scheduling processing time and computational complexity. A scheduling technique that employs FD-PF and FD-M-LWDF alternately in adjacent TTIs is proposed

FD-PF packet scheduling algorithm selects a user that maximizes $\mu_{i,j}(t)$ on PRB $j$ at TTI $i$:

$$\mu_{i,j}(t) = \frac{r_{i,j}(t)}{R_i(t)} \tag{11}$$

$$R_i(t+1) = \left(1 - \frac{1}{t_c}\right) R_i(t) + \frac{1}{t_c} * rtot_i(t+1) \tag{12}$$

$$rtot_i(t+1) = \sum_{j=1}^{RB_{max}} I_{i,j}(t+1) * r_{i,j}(t+1)$$

$$I_{i,j}(t+1) = \begin{cases} 1 & \text{if packets of user } i \text{ are scheduled on PRB } j \text{ at TTI } t+1 \\ 0 & \text{if packets of user } i \text{ are not scheduled on PRB } j \text{ at TTI } t+1 \end{cases} \tag{13}$$

where $\mu_{i,j}(t)$ is the priority of user *i* on PRB *j* at scheduling interval *t*, $r_{i,j}(t)$ is the instantaneous data rate of user *i* on PRB *j* at scheduling interval *t*, $R_i(t)$ is the average throughput of user *i* at scheduling interval *t*, $I_i(t+1)$ is the indicator function of the event that packets of user *i* are selected for transmission at scheduling interval *t+1* and $t_c$ is a time constant. The design objective of FD-PF algorithm is to maximize the long term throughput of the user whose current achievable data rate is better compared to the average throughput.

FD- M-LWDF packet scheduling algorithm selects a user that maximizes $\mu_{i,j}(t)$ in Equation (14) to receive its packet in each scheduling interval.

$$\mu_{i,j}(t) = \alpha_i \quad W_i(t) \quad \frac{r_{i,j}(t)}{R_i(t)} \tag{14}$$

$$\alpha_i = -\frac{(\log \delta_i)}{T_i} \tag{15}$$

where $\mu_{i,j}(t)$ is the priority of user *i* on PRB *j* at scheduling interval *t* and $\alpha_i$ is the QoS requirement of user *i*, $W_i(t)$ is the delay of the HoL packet of user *i* at scheduling interval *t*. Moreover, $r_{i,j}(t)$ represents the instantaneous data rate of user *i* on PRB *j* at scheduling interval *t* and $R_i(t)$ is the average throughput of user *i* at scheduling interval *t*. In addition, $\delta_i$ indicates the service-dependent PLR threshold of user *i* and $T_i$ is the service-dependent buffer delay threshold of user *i*. FD-M-LWDF algorithm was proposed to support QoS of multiple real time data users sharing a wireless channel. This algorithm is based on an overall consideration of packet delay information, the instantaneous data rate and average throughput and supposed to have a reasonably good overall performance.

The TD grouping technique divides the whole scheduling TTI into two groups: odd numbered TTI group and even numbered TTI group. FD-PF and FD-M-LWDF algorithm are employed in the two groups alternately. Each packet scheduling interval only employs one single scheduling algorithm which reduces the computational complexity compared with other joint TD-FD packet scheduling algorithms that prioritize users twice. Due to the fact that this algorithm uses two different scheduling algorithms in long-term, it is supposed to have reasonable performance improvement.

## 5. THE JOINT FD PREDICTION AND TD GROUPING PACKET SCHEDULING ALGORITHM

The main principle of the proposed algorithm is: firstly, at each TTI, the optimal estimated CQI value of each user is predicted by Kalman filter on each PRB, moreover the highest supported data rate of each user on each PRB can be estimated, and then the scheduling algorithm employed at this TTI is decided by the TD grouping technique which makes the scheduling decision based on the estimated CQIs.

Based on the above analysis, the proposed packet scheduling algorithm is described in Figure 2.

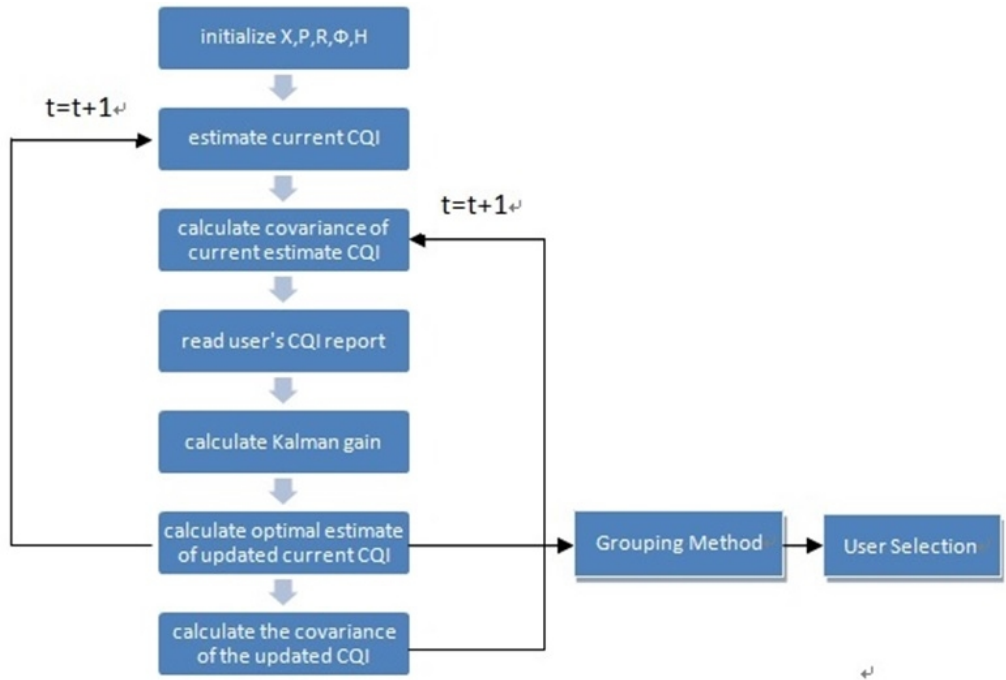

Figure 2: Flow chart of the proposed algorithm

## 6. SIMULATION RESULTS

The metrics such as system throughput and service packet loss ratio (PLR) are used in this paper to evaluate packet scheduling performance. System throughput is defined as the total size of successfully received packets (in bits) at all users in the downlink divided by the simulation time. The system throughput has following mathematical expression:

$$system\ throughput = \frac{1}{T} \sum_{i=1}^{N} \sum_{t=1}^{T} prx_i(t) \qquad (16)$$

where $prx_i(t)$ is the total size of correctly received packets (in bits) of user $i$ at time $t$, $T$ is the total simulation time and $N$ is the total number of users.

Service PLR is an important quality of service (QoS) metric, to satisfy QoS requirements, it has to be kept below a threshold throughout a session. For real time applications, the threshold is $10^{-6}$ (i.e. QoS constraints of the GBR (guaranteed bit rate) services). The service PLR is defined as the ratio of the total size of discarded packets of a service to the total size of all packets of a service that have resided in the eNB buffer. The mathematical expression for the service PLR is given below:

$$PLR = \frac{\sum_{i=1}^{N} \sum_{t=1}^{T} pdiscard_i(t)}{\sum_{i=1}^{N} \sum_{t=1}^{T} psize_i(t)} \qquad (17)$$

where PLR is the packet loss rate and $pdiscard_i(t)$ represent the total size of discarded packets (in bits) of user $i$ of a service at time $t$. $psize_i(t)$ is the total size of all packets (in bits) that have arrived into the eNB buffer of user $i$ of a service at time $t$, $N$ is the total number of users and $T$ is the total simulation time.

The simulation environment settings described in Table 1 is simulated. The FD-PF and FD-M-LWDF schedulers mainly serve as reference schedulers. Figure 3 and Figure 4 show the PLR of the FD-PF, FD-M-LWDF and the proposed algorithm with increasing number of users in the system (system capacity) under perfect CQI and imperfect CQI, respectively. It was observed in Figure 3 that the FD-PF algorithm has the worst PLR followed by the proposed algorithm and the FD-M-LWDF algorithm. To satisfy the QoS of GBR services, the maximum number of users that FD-PF algorithm can support is less than 30. Both of the proposed algorithm and the FD-M-LWDF algorithm can support 30 users. The use of FD-M-LWDF scheduler in the TD grouping based algorithm has significantly reduced the PLR compared with the FD-PF only scheduling and achieved the same system capacity with FD-M-LWDF algorithm, it is due to the fact that they consider packet delay information when making scheduling decision. Table 2 illustrates the proposed algorithm has 116.94% and 6.31% improvement in service PLR over FD-PF and FD-M-LWDF algorithm at 70 users, respectively.

Figure 4 shows that the proposed algorithm outperforms the other two algorithms and has the lowest PLR when the reported CQIs have 3 ms delay and become unavailable at a regular interval (10 ms). This is due to the use of the Kalman filter based CQI estimator. Packet scheduling decisions are made based on the estimated CQI instead of the erroneous CQI. It can be seen in the figure that both FD-PF and FD-M-LWDF scheduler have a substantial increase in PLR, which leads to a degradation in system capacities that fall below 30 users. This illustrates that the FD-PF and FD-M-LWDF scheduler are not tolerant of the imperfect CQI.

The system throughput of the three algorithms with increasing system capacity is shown in Figure 5 and Figure 6. A perfect CQI scenario is depicted in Figure 5 while Figure 6 demonstrates a practical system model with imperfect CQI. It can be seen from Figure 5 that the differences of the three algorithms are not obvious until the system capacity is above 50 users. FD-PF has the worst throughput performance while the proposed algorithm and the FD-M-LWDF algorithm have similar system throughput.

Similar to Figure 4, Figure 6 shows the proposed algorithm outperforms the two reference algorithm when a practical channel condition with imperfect CQI is considered. An improvement of 8.56% and 4.42% in system throughput over FD-PF and FD-M-LWDF algorithm with 70 users is illustrated in Table 3, respectively.

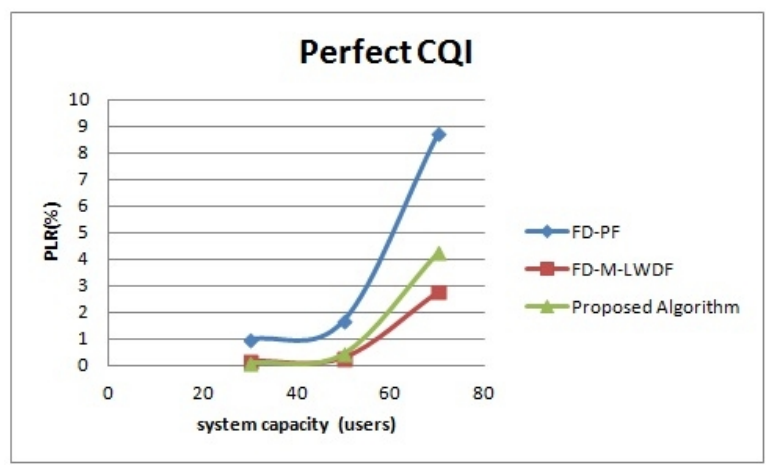

Figure 3 PLR comparison with perfect CQI

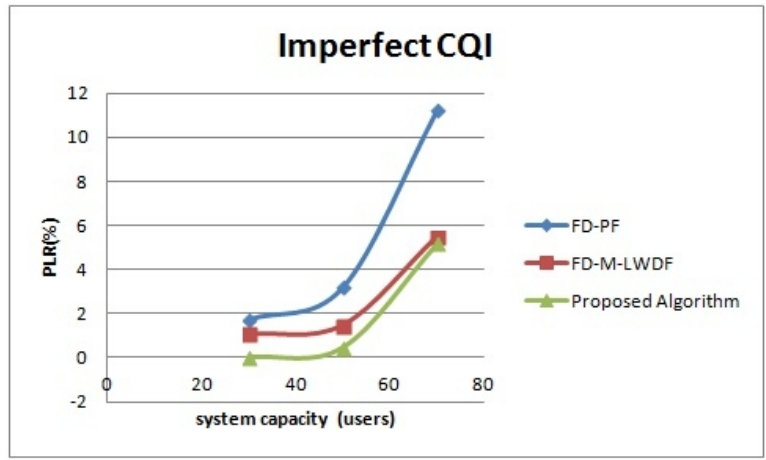

Figure 4 PLR comparison with imperfect CQI

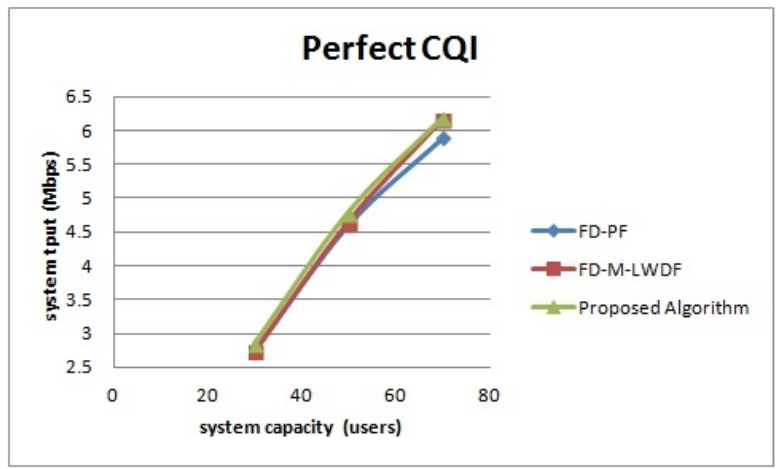

Figure 5 System throughput comparison with perfect CQI

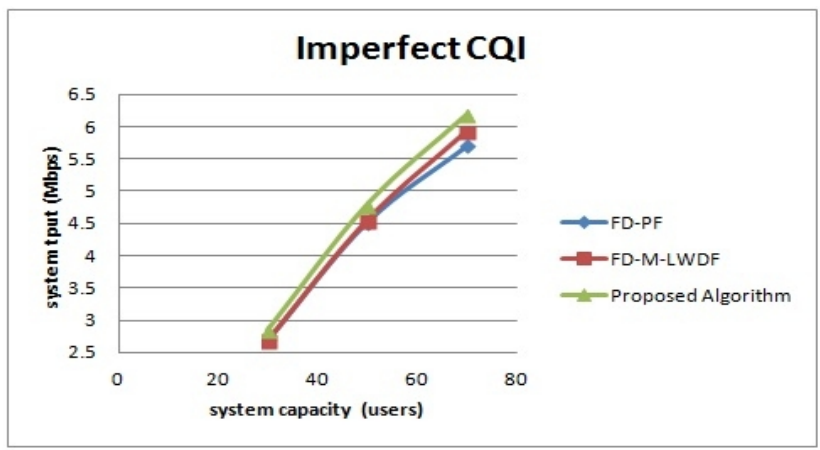

Figure 6 System throughput comparison with imperfect CQI

Table 2 PLR at 70 users

|  | PLR | Performance improvement (%) |
|---|---|---|
| FD-PF algorithm | 11.202 | 116.94 |
| FD-M-LWDF algorithm | 5.4898 | 6.31 |
| Proposed algorithm | 5.1637 | - |

Table 3 System throughput at 70 users

|  | System throughput (Mbps) | Performance improvement (%) |
|---|---|---|
| FD-PF algorithm | 5.6921 | 8.56 |
| FD-M-LWDF algorithm | 5.9174 | 4.42 |
| Proposed algorithm | 6.1792 | - |

## 7. CONCLUSIONS

This paper investigates the performance of three packet scheduling algorithms in the ideal and practical downlink 3GPP LTE system, respectively. A novel packet scheduling algorithm based on FD prediction and TD grouping technique for real time applications in the downlink LTE system is proposed in this paper. The simulation results show that the proposed algorithm outperforms FD-PF and M-LWDF algorithm in terms of service PLR and system throughput when the CQI reports has delay and errors. This simple and efficient algorithm which is robust to simultaneous multiple channel defects are proved to satisfy QoS requirements for real time users.

**Authors**

**Yongixn Wang** is currently a Master of Engineering by research Candidate in the Faculty of Engineering and Information Technology, University of Technology, Sydney (UTS), Australia. She received a Master of Engineering Degree in Software Engineering from Huazhong University of Science and Technology (2007) and Bachelor of Science Degree in Electrical and Electronic Engineering from Wuhan University of Science and Technology (2004). Her current research interests focus on packet scheduling in radio resource management for the future wireless networks

**Kumbesan Sandrasegaran** (Sandy) holds a PhD in Electrical Engineering from McGill University (Canada) (1994), a Master of Science Degree in Telecommunication Engineering from Essex University (UK) (1988) and a Bachelor of Science (Honours) Degree in Electrical Engineering (First Class) (UZ) (1985). He was a recipient of the Canadian Commonwealth Fellowship (1990-1994) and British Council Scholarship (1987-1988). He is a Professional Engineer (Pr. Eng) and has more than 20 years experience working either as a practitioner, researcher, consultant and educator in telecommunication networks. During this time, he has focused on the planning, modelling, simulation, optimisation, security, and management of telecommunication networks.

**Xinning Zhu** is currently a visiting scholar in the Faculty of Engineering and Information Technology, University of Technology, Sydney (UTS), Australia. She received her PhD, MS and BS degree in Communication and Information System from Beijing University of Posts and Telecommunications (BUPT) in 2010, 1995 and 1992. She is an associate professor at School of Information and Communication Engineering, BUPT, China. Her current research interests focus on interference management and mobility management in radio resource management for heterogeneous networks

**Jingjing Fei** is currently a Master of Engineering by research Candidate in the Computer Science and Engineering, University of New South Wales (UNSW), Australia. He received a Master of Engineering Degree in Software Engineering from Huazhong University of Science and Technology (2007) and Bachelor of Science Degree in Computer Science and Technology from Wuhan University of Technology (2004). His current research interests focus on wireless sensor networks.

**Xiaoying Kong** has broad interests in control engineering and software engineering. Her recent research work has been in GPS, inertial navigation systems, data fusion, robotics, sensor networks, Internet monitoring systems, agile software development methodologies, web technologies, web design methodologies, web architecture framework, analytical models of software development, and value-based software engineering. She has many years work experience in aeronautical industry, semiconductor industry, and software industry.

**Cheng-Chung** Lin is currently a PhD Candidate in the Faculty of Engineering and Information Technology, University of Technology, Sydney (UTS), Australia. He received a graduate certificate (GradCert) in advanced computing (2006) and a Master of Information Technology in internetworking in School of Computer Science and Engineering from University of New South Wales, Australia (2007). His current research interests focus on handover and packet scheduling in radio resource management for the future wireless networks